\documentclass{article}
\usepackage[T1]{fontenc} 
\usepackage[utf8]{inputenc} 
\usepackage{ismir,amsmath,cite,url}

\usepackage{graphicx}
\usepackage{color}
\usepackage[table]{xcolor}
\usepackage{times}
\usepackage{soul}
\usepackage{url}

\usepackage[utf8]{inputenc}
\usepackage[small]{caption}
\usepackage{subcaption}
\usepackage{graphicx}
\usepackage{amsmath}
\usepackage{amsthm}
\usepackage{booktabs}
\usepackage{enumitem}
\usepackage{multirow}
\usepackage{algorithm}
\usepackage{algorithmic}

\usepackage{lineno}
\title{Symphony Generation with Permutation Invariant Language Model}

\multauthor
{Jiafeng Liu$^{1\star}$\thanks{$^{\star}$ The authors contributed equally to this work.}\hspace{1cm} Yuanliang Dong$^{1
\star}$ \hspace{1cm} Zehua Cheng$^2$} { \bfseries Xinran Zhang$^{1}$ \hspace{1cm} {Xiaobing Li$^{1}$ \hspace{1cm} Feng Yu$^{1}$ \hspace{1cm} Maosong Sun$^{1,3\dag}$  \thanks{$^{\dag}$ Corresponding author. Email: sms@tsinghua.edu.cn}           }\\
 $^1$ Department of Music AI and Music Information Technology, Central Conservatory of Music\\
$^2$ Department of Computer Science, University of Oxford\\
$^3$ Department of Computer Science and Technology , Tsinghua University\\ 
{\tt\small \{jiafeng.liu, gunterdong\}@mail.ccom.edu.cn}\\
}

\def\authorname{J. Liu, Y. Dong, Z. Cheng, X. Zhang, X. Li, F. Yu, M. Sun}
\usepackage[bookmarks=false,pdfauthor={\authorname},pdfsubject={\papersubject},hidelinks]{hyperref}
\begin{document}

\maketitle
\begin{abstract}
In this work, we propose a permutation invariant language model, SymphonyNet, as a solution for symbolic symphony music generation. We propose a novel Multi-track Multi-instrument Repeatable (MMR) representation for symphonic music and model the music sequence using a Transformer-based auto-regressive language model with specific 3-D positional embedding. To overcome length overflow when modeling extra-long symphony tokens, we also propose a modified Byte Pair Encoding algorithm (Music BPE) for music tokens and introduce a novel linear transformer decoder architecture as a backbone. Meanwhile, we train the decoder to learn automatic orchestration as a joint task by masking instrument information from the input. We also introduce a large-scale symbolic symphony dataset for the advance of symphony generation research. Empirical results show that the proposed approach can generate coherent, novel, complex and harmonious symphony as a pioneer solution for multi-track multi-instrument symbolic music generation.
\end{abstract}
\section{Introduction}
\label{INTRO}
Symphony is one of the most complex and brilliant musical composition forms in human history, where many instruments are intertwined to express rich human emotions. The past decade has seen the rapid development and tremendous success of the symbolic music generation in both research and industrial field~\cite{deepbach,midinet,musicvae}. Most current works follow conventional text modeling and generation method by applying language model to sequences of symbolic musical events ~\cite{mt,popmag,cwt}. However, symphony modeling and generation still constitutes in itself a considerable challenge since symphony
music sequences differ from text sequences in various aspects.

\begin{figure}[t]
  \centering
  \centerline{\includegraphics[width=1.1\columnwidth]{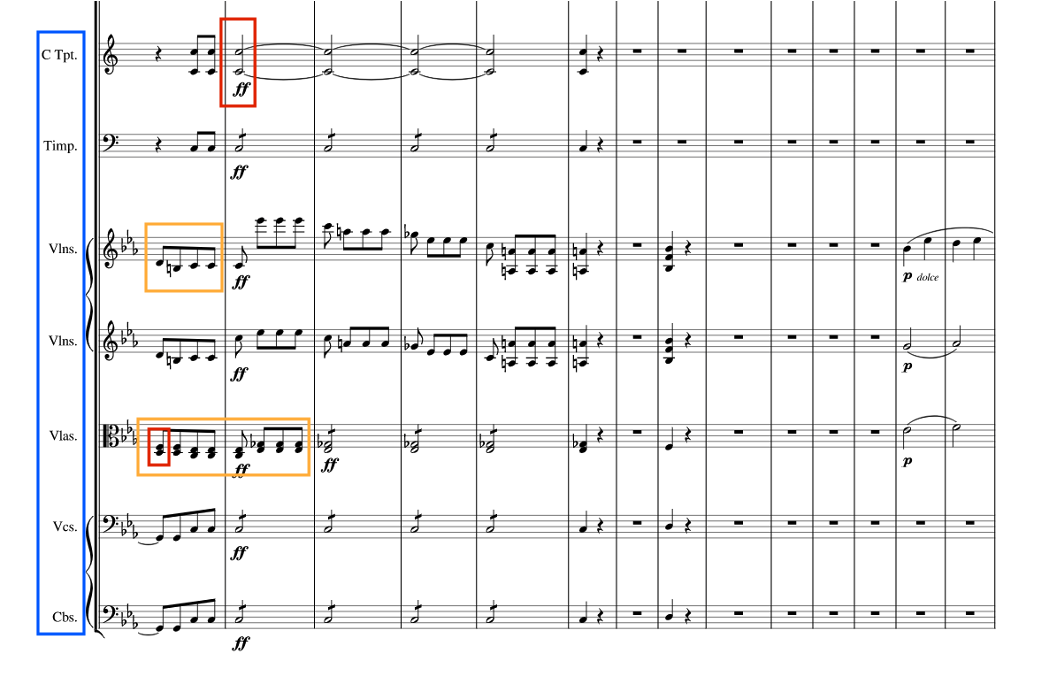}}
  \caption{A simple example of Multi Instruments \& Multi Tracks \& Repeat Instruments symphony score.}
  \label{fig:score}
\end{figure}

Natural language could be modeled as a purely linear sequence constructed strictly by a sequential order of words. Symphony scores, on the other hand, are usually viewed as two-dimensional symbolic sequences in which many notes can be played concurrently. Notes in a symphony score are \textbf{semi-permutation invariant}. More specifically, as shown in Fig.~\ref{fig:score}, the blue box indicates the musical instrument tracks, and the corresponding staves on the right side are permutation invariant. Similarly, the notes inside the red box are also permutation invariant. In contrast, notes in the upper yellow box are permutation variant since each note is played sequentially.  Changes in the order of notes will impair the music itself. The yellow box at the bottom is a more complicated situation: a permutation variant note sequence in general containing permutation invariant notes. Simply flattening the score into a 1-D text-like sequence may damage the local structure of music~\cite{musebert}. To address this problem, we propose the Multi-track Multi-instrument Repeatable (MMR) representation with particular 3-D positional embedding in Section~\ref{MMR Representation} which fully considers the properties of semi-permutation invariance in symbolic music scores.

Moreover, when comparing music scores with text, conventionally notes could be considered as characters, while intervals or chords are comparable to words. Modeling musical events at note level is a common practice~\cite{popmt,popmag,cwt,musicbert}. However, this may be confronted with similar problems in char-level text generation, such as extremely long sequences and less meaningful individual tokens. Word-level tokenization suffers from large vocabulary size and out of vocabulary (OOV) problems. Byte Pair Encoding (BPE) \cite{bpe-nlp,bpe} subword tokenization is a tradeoff between word-level and character-level tokenization. Inspired by BPE, we propose the Music BPE algorithm in Section~\ref{Music BPE}, which could automatically aggregate notes to intervals and chords as subwords without a pre-defined vocabulary and construct music sequences with richer semantics.

\label{para:intro-note}

Generating symphony music with proper instruments for different tracks is another challenging task.  Recent work like Arranger~\cite{arranger} focuses on instrumentation by learning to separate parts from the mixture in symbolic multi-track music. However, it does not incorporate music generation task. In this paper, we present a unique linear transformer decoder architecture for instrument classification with joint-task training, which allows the model to learn auto-orchestration rather than relying on instrument information as an pre-defined input source. ~\cite{musenet,popmag,lakhnes,mmm}.

The contributions of this paper are presented as below:
\begin{itemize}
    \item We propose a novel Multi-track Multi-instrument Repeatable (MMR) representation for symphony music, including particular 3-D positional embedding designed to address the  semi-permutation invariant challenge in symphony generation. Our method is also compatible with all existing symbolic music ensembles, including but not limited to piano solo, quartet and pop band music.
  
    \item We propose a novel algorithm, Music BPE, to model the symbolic music at subword-level. Furthermore, we found that our Music BPE algorithm could aggregate notes to intervals and chords, which are consistent with common chords summarized by human musicians. 
    \item We introduce SymphonyNet, a novel music generation model with joint-task training for instrument classification based on our proposed MMR representation and Music BPE. The model can learn the proper orchestration according to the distribution of the notes.
    
    \item  We collect a symphony MIDI dataset, consisting of $46,359$ high-quality MIDI files with multiple instruments and tracks to advance researches on symphony generation with deep learning.
   
\end{itemize}
\section{Related Work}
\begin{table*}[htbp]
\renewcommand{\arraystretch}{1.2}
\small
  \centering
    \begin{tabular}{c|ccccc}
    \textbf{Name} & \textbf{Time Unit} & \textbf{Representation} & \textbf{Backbone Model} & \textbf{Type of Music} & \textbf{Instruments}\\
    \toprule
    DeepBach \cite{deepbach} & Fixed-length grid & N/A   & Bi-directional  RNN & Chorales & Fixed Ensemble\\
    MuseGAN \cite{musegan} & Fixed-length grid & N/A   & GAN   & Multi-track & Fixed Ensemble\\
    Music Transfor. \cite{mt}    & MIDI-event timeshift & N/A & Vanilla Transformer & Piano & N/A\\
    Pop MT \cite{popmt} & Beat and note duration & REMI  & Transformer-XL & Piano  & N/A\\
    CWT \cite{cwt}   &  Beat and note duration & Compound Word & Linear Transformer & Piano  & N/A\\
    Musenet \cite{musenet} & MIDI-event timeshift & Token Aggregation & GPT-2 & Multi-track & Not Repeatable\\
    PopMAG \cite{popmag} &  Beat and note duration  & MuMIDI & Transformer-XL & Multi-track  & Not Repeatable\\
    LakhNES \cite{lakhnes} & MIDI-event timeshift & Token Aggregation & Transformer-XL & Multi-track & Fixed Ensemble\\
    MMM \cite{mmm}   & MIDI-event timeshift & Hierarchical & GPT-2 & Multi-track &  Repeatable\\
    This work &  Beat and note duration  & MMR   & Linear Transformer & Multi-track & Repeatable\\
    \midrule
    PiRhDy \cite{pirhdy} & Fixed-length grid & Fusion module & RNN with attention &  Multi-track  & Not Repeatable\\
    MusicBert \cite{musicbert} &  Beat and note duration  & OctupleMIDI & Roberta & Multi-track & Not Repeatable\\
    \bottomrule
    \end{tabular}%
  \caption{An overview of time unit, representation, backbone model and music type in existing works, above for generation works and below for understanding works. }
  \label{tab:previous-work}%
\end{table*}%

We organize some existing works in Table \ref{tab:previous-work} in terms of five aspects of symbolic music modeling: time unit, representation method, backbone model,  music type and the ability to model music with repeat instruments. Generation works are presented above and understanding works are presented below. Pianoroll, MIDI event timeshift, and Beat-based onset and duration are the mainstream time units in music generation and understanding tasks. However, Pianoroll divides music into fixed-length grids, and MIDI format provides overprecise timeshift events, both suffering from sparsity problems, which raises another handicap for applying deep learning models in this multi-track generation.
Pop Music Transformer~\cite{popmt} is the first attempt to introduced the beat-based REMI representation in music generation. It supports variable-length duration of notes, which is more musically inspired. Compound Word~\cite{cwt}, derived from REMI representation, classifies the sequence of REMI into note-related or metric-related events, which are then aggregated, greatly decreasing the sequence length.. This has engendered a new trend of beat-based symbolic music generation. 

Language models are now prevalent in natural language processing tasks~\cite{brown2020language}. However, applying language models to the creation of multi-track music remains challenging.
MuMIDI~\cite{popmag} and OctupleMIDI~\cite{musicbert} models multiple attributes of one note in one sequence step and also incorporates instrument tokens for multi-track representation. However, if one musical piece contains more than one track for the same instrument, their representation could not distinguish them in different tracks. MMM~\cite{mmm} introduced a MIDI-event-like representation, creating a time-ordered sequence of musical events for each track and concatenating several tracks into a single sequence. However, MMM adopts time-delta tokens and fixed positional encoding which weakens the note-level correlation and structure between tracks. MuseBert~\cite{musebert} proposes a permutation invariant bert-like language model with generalized relative position encoding (RPE) which, however, is not compatible with multi-track music generation.

Though various symbolic music representation strategies have been proposed, few are compatible with multi-track music with repeatable instruments or tracks, such as the symphony. Furthermore, permutation invariance of music, as is discussed in Section~\ref{INTRO}, has scarcely been considered. To our knowledge, this work proposes the first representation and tokenization method to encode music with multiple repeatable instruments and multiple repeatable tracks and designs a universal and effective strategy for generating symphony music with permutation invariant language model.
\section{Multi-track Multi-instrument Repeatable Representation}\label{MMR Representation}
To further analyze the symphony generation task, it is crucial to understand the difference between the symphony format and other genres of music.
\begin{itemize}[leftmargin=*]
 \item \textbf{Single Instrument in Single Track}. No more than one note is played at any timestep by one instrument. Also called monophonic music. e.g., flute.
 \item \textbf{Multi Instruments \& Each in Single Track}. Only one note for each instrument is played at any timestep. e.g., quartet singing.
 \item \textbf{Single Instrument in Multi Tracks}. There are multiple notes played in each timestep while only one instrument. e.g., piano.
 \item \textbf{Multi Instruments \& Multi Tracks \& No Repeat Instrument}. There are multiple notes played in each timestep. No constraint on the number of instruments and all instruments are unique. e.g., classical pop band with only drum, electric guitar and bass.
 \item \textbf{Multi Instruments \& Multi Tracks \& Repeat Instruments}. Instruments are not unique and multiple same instruments can play different notes on different tracks, e.g., symphony.
\end{itemize}

For the last case, it's a common practice to merge the same instruments into a single track in previous works. However, it may damage the intrinsic structure of symphony music. For example, this may cause a violin to play polyphonic notes, or even intermingle multiple melody lines. Our proposed Multi-track Multi-instrument Repeatable (MMR) representation models repeated instruments separately, which could capture more heuristic musical information within a single track. Since our MMR representation is aimed at symphony modeling, it is also compatible with all existing music ensembles.

\begin{figure*}[t]
  \centering
  \centerline{\includegraphics[width=2.1\columnwidth]{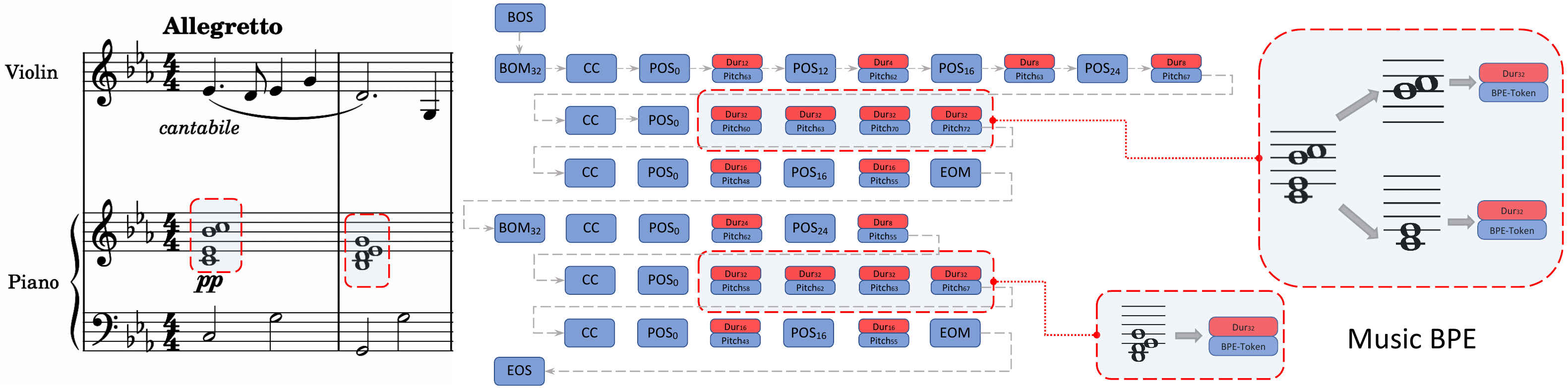}}
  \caption{An example of MMR representation and illustration of Music BPE process}
  \label{fig:mmr}
\end{figure*}

\subsection{Structural and Controlling Token}
We consider that special tokens perform two primary functions in a symphony music generation task: 1) To represent the musical structures of notes. 2) To control the model output during the inference phase.

\textbf{Score} We use a pair of $[BOS]$ and $[EOS]$ tokens to designate the beginning and end of a symphony score.

\textbf{Measure} Different from~\cite{popmag,popmt}, we ascribe a pair of $[BOM_i]$ and $[EOM]$ to indicate the beginning and end of a measure, the character $i$ to represent the total length of the current measure. The length of a measure is calculated by time signature, and we choose 32\textsuperscript{th} note as the smallest unit of time. For example, a 4/4 time signature indicates four quarter notes length per measure, which is equal to the length of thirty-two 32\textsuperscript{th} notes. In that case, character $i$ equals 32, and the measure beginning token is $[BOM_{32}]$ 

\textbf{Chord} The chord token is a valuable indicator of how generally notes are arranged in the current measure. We pre-define $132$ common types of chord token and pre-compute chord tokens with a rule-based algorithm proposed in~\cite{cwt}, such as C major seventh chord marked as token $[C_{maj7}]$.

\textbf{Track} Unlike any previous works, we do not explicitly encode the track and instrument transformation in a single token. A change track token $[CC]$ only signifies the start of a new track for the latter controlling purpose. Section \ref{symphonyNet} will further explore the traits of tracks and instruments and the approaches of differentiating tracks.

\textbf{Position} A position token stands for the \emph{onset} of a note within the measure, represented by the token $[POS_j]$. The following event tokens are controlled by the current position token until another position token shows up. The character $j$ means the number of the current unit of time position. For example, a $[POS_{48}]$  indicates the 48\textsuperscript{th} unit time position.

To summarize, structural and controlling tokens are designed to specify the general time-spatial features of notes, such as the time a note is to be played and the track it locates. In this work, these tokens are mandated with a sequential order, as a \emph{measure} token shall be followed by a \emph{chord} token, which altogether represents in a explicit way the measure order as shown in Fig.~\ref{fig:model}.
\subsection{Note-Related Tokens}
A note in music scores could be defined in five attributes: pitch, duration, onset, track and instrument. Pitch and duration are content-related and the others are position-related. The latter will be discussed in Section~\ref{symphonyNet}. To avoid the long-tail problem, we regard pitch and duration to be distinct note properties and construct two separate vocabularies for model input. Then we aggregate note pitches with identical duration and onset by our proposed Music BPE algorithm, as will be described in the next section.

\section{Music Byte Pair Encoding} 
\label{Music BPE}
As shown in Fig.~\ref{fig:mmr}, a complex chord is constructed by several notes at the same position in a measure, which can be deconstructed into two common and simple intervals. Unlike natural language, notes played at the same position are permutation invariant. Changing the order of notes in a chord does not affect its sound or meaning. For instance, a Chord $C$ consists of ($C4$, $E4$, $G4$), which is equal to ($G4$, $C4$, $E4$). This intrinsic property may conflict with the typical natural language processing job, imposing a new constraint on the use of conventional tokenization methods such as standard BPE.

In this work, we propose a novel encoding approach, Music Byte Pair Encoding (Music BPE), for multi-track symbolic music sequence tokenization and preprocessing to exploit the semantics of music events and minimise the length of the input context from a representation standpoint. Different from the original BPE algorithm, our proposed Music BPE is based on \textbf{concurrence} of notes rather than \textbf{adjacency} of characters.

Our implementation of Music BPE is shown in Algorithm~\ref{alg:pimbpe}. As is mentioned in Section~\ref{para:intro-note}, a note has five attributes: \emph{pi}tch, \emph{du}ration, \emph{po}sition, \emph{tr}ack and instrument, while the instrument depends utterly on track within the same measure. Formally, in a piece of symbolic music, we define a maximum set of two or more notes 
\[ \{(pi, du, po, tr) \mid \text{where } du,po,tr \text{ is constant} \} \]
as a \emph{mulpi} (multiple pitches), i.e., a maximum set of notes that have the same duration at the same global position and within the same track, equivalent to a "word" in the BPE algorithm.

\begin{algorithm}[ht]
\caption{Music BPE}
\label{alg:pimbpe}
\textbf{Input}: A multi-set of \emph{mulpies} $B$  \\
\textbf{Parameter}: desired dictionary size $n$\\
\textbf{Output}: Merged dictionary $V$ 
\begin{algorithmic}[1] 
\STATE Let $V = \{ \{ p \} \mid p\in[0, 128) \}$.
\STATE Let $C$ be an empty multi-set
\FORALL{$mulpi \in B$}
\STATE $mulpi \gets \{ \{ p \} \mid p\in mulpi \}$
\FORALL{$\{P_1,P_2\} \subseteq mulpi$}
\STATE Insert $(P_1,P_2)$ into $C$.
\ENDFOR
\ENDFOR
\WHILE{$|V| < n$} 
\STATE Let $(P_1,P_2)$ be the most frequent pair in $C$.
\STATE $V \gets V \cup \{P_1 \cup P_2\}$ 
\FORALL{$mulpi \in B$}
\IF{$\{P_1, P_2\} \subseteq mulpi$}
\FORALL{$Q \in mulpi - \{P_1, P_2\}$}
\STATE Delete $(Q,P_1),(Q,P_2)$ from $C$.
\STATE Insert $(Q, P_1 \cup P_2)$ into $C$.
\ENDFOR
\STATE $mulpi \gets (mulpi - \{P_1, P_2\}) \cup \{P_1 \cup P_2\}$
\ENDIF
\ENDFOR
\ENDWHILE
\STATE \textbf{return} $V$
\end{algorithmic}
\end{algorithm}
We collect notes with the same global position and the same duration in the same track from each music piece to construct a bag of \emph{mulpies}. The vocabulary list is initialized with $128$ MIDI pitches, where each token represents a pitch-set containing a single pitch. Every time we locate all concurrent pairs of tokens in the bag of \emph{mulpies}, merge the most frequent pair ('$P_1$', '$P_2$') into a new symbol $P$ and replace the pair with the new symbol in each \emph{mulpi} until the vocabulary size reaches the maximum limit. A further discussion on the results of the Music BPE algorithm and its effectiveness on our symphony dataset will be presented in Section~\ref{symphonyNet}.
\nopagebreak
\section{SymphonyNet Details} 
\label{symphonyNet}

\subsection{The 3-D Positional Embedding}
\begin{figure*}[ht]
  \centering
  \centerline{\includegraphics[width=2.1\columnwidth]{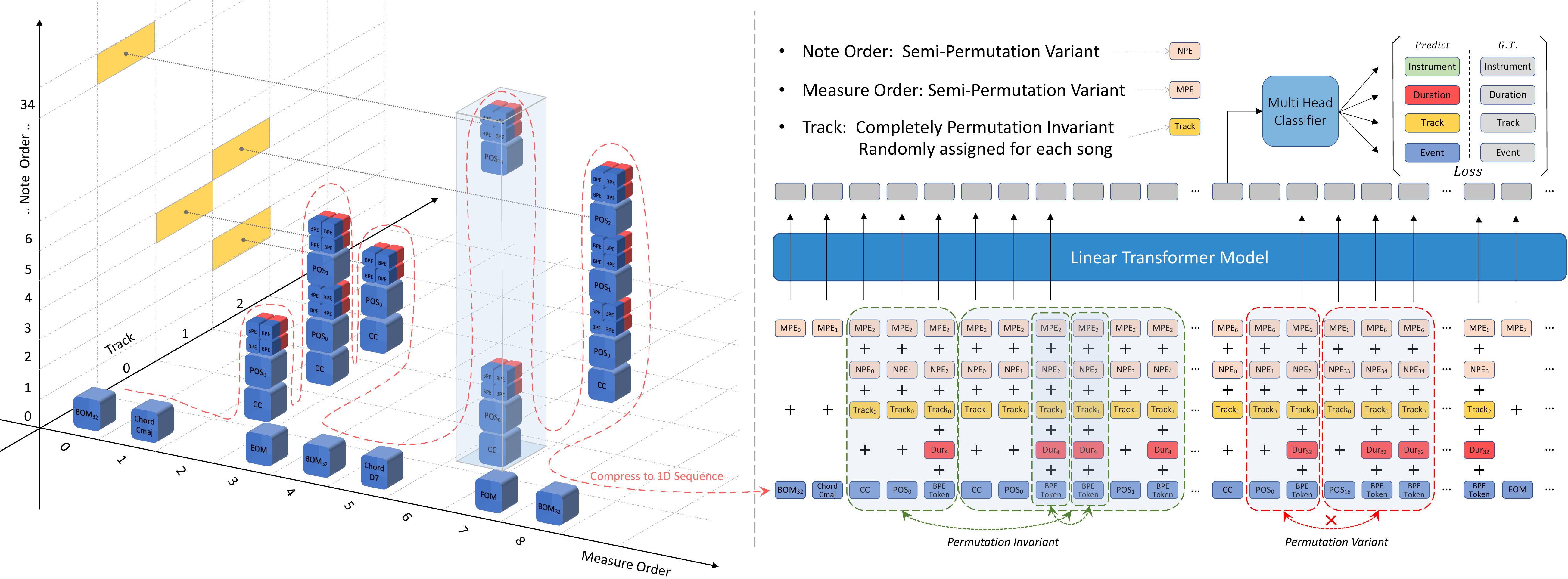}}
  \caption{A illustration of the spatial and structural attributes of MMR sequence (left) and the way it is compressed and organized as model input (right).}
  \label{fig:model}
\end{figure*}

Transformer~\cite{vaswani2017attention} is the most used backbone for language model, which is designed \emph{permutation invariant}: if the positional encoding is not added, disrupting the order of the inputs will yield the same output, for transformer model treats inputs as a \emph{set} during self-attention. Therefore, considering this property of Transformer, we design a particular 3-D positional embedding to represent such a semi-permutation invariant feature as shown in Fig.~\ref{fig:model}. Event tokens follow a semi-permutation variant order on both the measure order and the note order axes. For example, notes played on the same position share the same note positional embedding. In contrast, the track axis is entirely permutation invariant since we only need track embeddings to differentiate tracks other than a sequential order. We use the red curves to illustrate the musical moving trajectory of event tokens to better understand how we compress the spatial and structural sequence of the event tokens into one dimension and send them to the model. At last, we add all constructed embeddings vertically as the model input. To address the extraordinarily long symbolic music sequences challenge, we employ the linear transformer~\cite{lineartrans} as the backbone of our model to satisfy the length constraint. The model follows a decoder-only fashion, and we design different feed-forward heads for four attributes of musical events, which are \textbf{Instrument, Track, Duration, and Event} tokens as shown in Fig.~\ref{fig:model}.

\subsection{Joint Task with Instrument Classification}

We mask instrument information for every input token at the input side, and anticipate that the model will learn instrumentation from the output side with instrument loss. This will turn a succession of simple, blank notes into a fully orchestrated piece of music, analogous to colouring a black-and-white painting. This design is motivated by two primary concerns. First, we investigate the possibility if other instrument may play a certain instrument's note track. Therefore, that is a case for the model to determine to what degree the instrument fits the track's notes and how instruments interact with one another across tracks. For instance, it is allowed to substitute the piano for the marimba in some musical compositions. The intrinsic nature of a pre-assigned instrument for notes reduces the diversity of training data.

\section{Experiments and Results} \label{Result}
This section introduces the novel symphony dataset we propose and presents two stages in the training process\footnote{Our code, demos, dataset and further analysis can be accessed at \href{https://symphonynet.github.io}{https://symphonynet.github.io}}. Secondly, we describe controllable methods to generate music under certain condition before  we provide findings from Music BPE and compare them with the specific musical knowledge. Lastly, a human evaluation result analysis and scoring on several excerpts generated by different models will be presented.

\subsection{Symphony Dataset}
To tackle the obstacles of the symphony generation research, we gather a big corpus of symphonic music from multiple online sites and conduct a extensive data cleaning. The average duration of the 46,359 MIDI files containing multiple instruments and tracks, mostly symphony, is 4.26 minutes. The collection contains more than 279 million notes and 567 million tokens in MMR forms. Our symphony dataset is, to the best of our knowledge, the first worldwide large-scale symbolic symphonic music dataset, which might serve as a foundation for future work in multi-track music production.

\begin{figure}[t]
  \centering
  \centerline{\includegraphics[width=1.05\columnwidth]{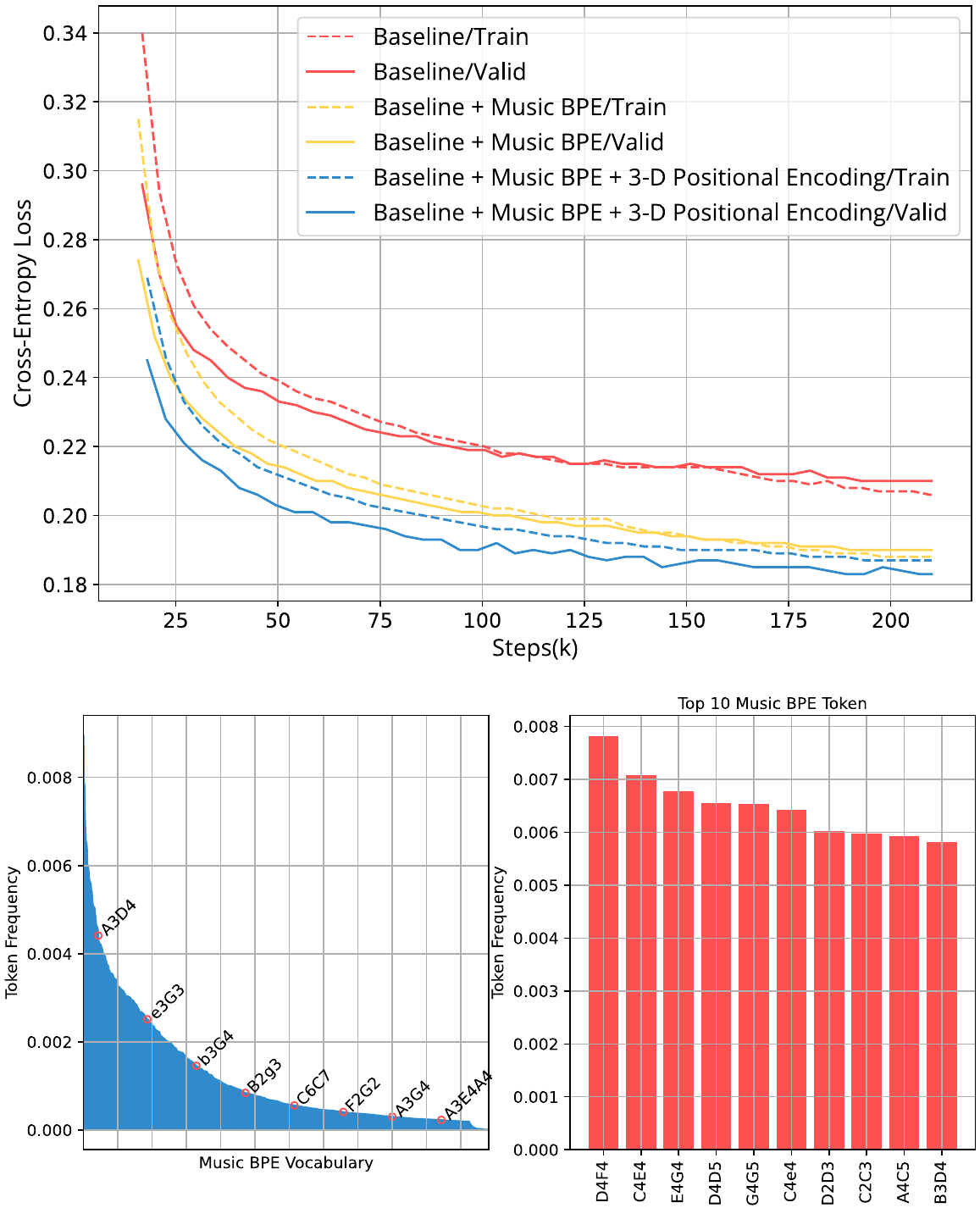}}
  \caption{The training and validation curves of different models and the Music BPE note aggregation results.}
  \label{fig:loss}
\end{figure}

\subsection{Training Details}
In our experiment, the model adopts $4096$ as the length of input sequence. We set the embedding size for event tokens, durations, instruments and 3-D positional embedding to $512$. The final size of event token vocabulary is assigned to $1000$ after running Music BPE algorithm and the vocabulary size of durations, instruments, 3-D positional embeddings are derived from the dataset. The linear transformer decoder contains $12$ self-attention layers and each layer consists of 16 attention heads. SymphonyNet is trained with eight 2080 Ti GPU and we use a batch size of $128$ and an AdamW~\cite{adamw} optimizer with a learning rate of $3 \times 10^{-4}$.

\subsection{Music BPE Result}
After constructing a vocabulary list of length $1,000$ with Music BPE algorithm, the top-5 merged pairs shown in Fig. \ref{fig:loss} with the highest frequency are: ($D4$, $F4$), ($C4$, $E4$), ($E4$, $G4$), ($D4$, $D5$), and ($G4$, $G5$), which are usual intervals occurring in symphony \emph{divisi} passages. After applying Music BPE on the whole music corpus, the average token length of a $mulpi$ shortens to half (from $2.28$ to $1.13$), also reducing the burden of modeling ultra-long symphony sequences.

\subsection{Ablation Study and Human Evaluations}
We train a linear transformer decoder model with the vanilla positional encoding of GPT-3~\cite{brown2020language} as a baseline. Then we train another model of the same architecture with the training data processed by the proposed Music BPE algorithm. Finally, we incorporate both 3-D positional embedding and Music BPE algorithm, which achieves the lowest training and validation loss after the same total training steps, as is shown in Fig. \ref{fig:loss}. The objective metric indicates that our permutation-invariant 3-D positional embedding and Music BPE algorithm could significantly improve  model performance and generalization ability.

Also, we perform a human evaluation to compare the quality of generated music excerpts from different models with human composition. Participants include 25 professional musicians and 25 non-musicians. Each participant receives 16 excerpts: four excerpts conditioned on a chord progression, four excerpts conditioned on a given 4-bar prime, and eight unconditioned excerpts. The music excerpts are rated in 5 dimensions: Coherence \textbf{(C)}, Diversity \textbf{(D)}, Harmoniousness \textbf{(H)}, Structureness \textbf{(S)}, Orchestration \textbf{(O)} and Overall Preference \textbf{(P)}, in a 5-point Likert scale. As shown in Table~\ref{tab:human listen_A}, the model with 3-D positional embedding and Music BPE beats most of the approaches. It is worth noting that excerpts generated by our models surpass the human compositions in the indicator of diversity marked by yellow color.

To further explore the model performance, we retrain SympohyNet on Lakh MIDI Dataset \cite{lakh} with the same backbone model architecture as MMM~\cite{mmm}, and carry out another human evaluation to compare with MMM. Each participant receives 10 excerpts: five generated unconditionally from MMM and the others generated unconditionally from retrained SymphonyNet. The results are presented in Table~\ref{tab:human listen_B}, which indicate that SymphonyNet surpass MMM in all indicators.  Overall, the human hearing test suggests that SymphonyNet can construct coherent, unique, complex, and harmonic symphonies.
\begin{table}
\small
\begin{subtable}{\linewidth}

\setlength{\tabcolsep}{4.25 pt}
\renewcommand{\arraystretch}{1.15}
\begin{tabular}{llcccccc}
 \hline
 & \textbf{Model} & \textbf{C} & \textbf{D} & \textbf{H} & \textbf{S} & \textbf{O} & \textbf{P} \\ 
  \hline
\multirow{4}{*}{Chord} & Baseline & 3.5 & 3.57 & 3.07 & \multicolumn{1}{r}{3.00} & 3.21 & 3.29 \\
 & BPE & 3.64 & 3.64 & 3.14 & \textbf{3.15} & 3.43 & 3.29 \\
 & 3D + BPE & \multicolumn{1}{r}{\textbf{3.71}} & \multicolumn{1}{r}{\cellcolor[HTML]{FFFC9E}\textbf{3.72}} & \multicolumn{1}{r}{\textbf{3.21}} & 3.07 & \multicolumn{1}{r}{\textbf{3.5}} & \multicolumn{1}{r}{\textbf{3.5}} \\ \cline{2-8} 
 & Human & 4.43 & 3.43 & 4.14 & 4.36 & 4.14 & 4.14 \\ 
  \hline
 &  &  &  &  &  &  &  \\
  \hline
\multirow{4}{*}{Prime} & Baseline & 3.79 & 2.79 & 3.21 & 3.43 & 3.36 & 3.36 \\
 & BPE & \multicolumn{1}{r}{\textbf{3.86}} & \textbf{3.5} & \textbf{3.5} & 3.5 & 3.64 & \textbf{3.86} \\
 & 3D + BPE & \textbf{3.86} & 3.14 & 3.43 & \textbf{3.57} & \textbf{3.93} & 3.64 \\
 \cline{2-8} 
 & Human & 4.36 & 3.57 & 4.36 & 4.00 & 4.36 & 4.36 \\
  \hline
 &  &  &  &  &  &  &  \\
  \hline
\multirow{4}{*}{Uncondi.} & Baseline & 3.52 & 3.46 & 3.04 & 3.07 & 3.11 & 3.07 \\
 & BPE & \textbf{3.79} & 3.64 & 3.25 & 3.11 & 3.25 & 3.29 \\
 & 3D + BPE & 3.53 & \cellcolor[HTML]{FFFC9E}\textbf{3.93} & \textbf{3.43} & \textbf{3.32} & \textbf{3.43} & \textbf{3.32} \\
 \cline{2-8} 
 & Human & 4.39 & 3.89 & 4.18 & 4.21 & 4.11 & 4.29 \\
  \hline

\end{tabular}
\caption{Trained on Symphony Dataset}
\label{tab:human listen_A}
\end{subtable}
\hfill
\hfill
\begin{subtable}{\linewidth}

\setlength{\tabcolsep}{5.6 pt}
\renewcommand{\arraystretch}{1.25}
\begin{tabular}{lcccccc}
 &  &  &  &  &  & \\
\hline
\textbf{Model} & \textbf{C} & \textbf{D} & \textbf{H} & \textbf{S} & \textbf{O} & \textbf{P} \\ \hline
\multirow{2}{*}{MMM} & 3.20 & 2.71 & 2.51 & 2.66 & 2.80 & 2.71 \\
 & ±0.13 & ±0.12 & ±0.13 & ±0.12 & ±0.13 & ±0.11 \\ \hline
\multirow{2}{*}{\textbf{Symph.}} & \textbf{3.33} & \textbf{2.89} & \textbf{2.76} & \textbf{2.69} & \textbf{2.99} & \textbf{2.87} \\
 & ±0.15 & ±0.13 & ±0.12 & ±0.13 & ±0.12 & ±0.13 \\ \hline
\end{tabular} \caption{Trained on Lakh MIDI Dataset} \label{tab:human listen_B}
\end{subtable}
 \caption{Human evaluation results from 25 musicians and 25 non-musicians, with mean opinion scores and 95 percent confidence intervals reported.}
 \label{tab:human listen}%
\end{table}

\section{Conclusion}
In this work, we illustrate the properties of multi-track and multi-instrument music, like symphony, and propose a novel MMR representation with 3-D positional embedding for modelling it. To tokenize the ultra-long symbolic music sequence at sub-word level, we propose the Music BPE algorithm. Besides, we design a joint task for the model to learn auto-orchestration. Human evaluation results show that our suggested technique produces competitive symphonic music when compared to human compositions. In the future, we will investigate modelling long-term musical structures, since complex music, such as symphonies, often consists of numerous parts or movements.
\section{Acknowledgements}
Thanks for the anonymous reviewers for their valuable comments. This work is supported by High-grade, Precision and Advanced Discipline Construction Project of Beijing Universities, Major Projects of National Social Science Fund of China (Grant No. 21ZD19), and Nation Culture and Tourism Technological Innovation Engineering Project.
\bibliography{ISMIRtemplate}

\begin{thebibliography}{10}
\providecommand{\url}[1]{#1}
\csname url@samestyle\endcsname
\providecommand{\newblock}{\relax}
\providecommand{\bibinfo}[2]{#2}
\providecommand{\BIBentrySTDinterwordspacing}{\spaceskip=0pt\relax}
\providecommand{\BIBentryALTinterwordstretchfactor}{4}
\providecommand{\BIBentryALTinterwordspacing}{\spaceskip=\fontdimen2\font plus
\BIBentryALTinterwordstretchfactor\fontdimen3\font minus
  \fontdimen4\font\relax}
\providecommand{\BIBforeignlanguage}[2]{{%
\expandafter\ifx\csname l@#1\endcsname\relax
\typeout{** WARNING: IEEEtran.bst: No hyphenation pattern has been}%
\typeout{** loaded for the language `#1'. Using the pattern for}%
\typeout{** the default language instead.}%
\else
\language=\csname l@#1\endcsname
\fi
#2}}
\providecommand{\BIBdecl}{\relax}
\BIBdecl

\bibitem{deepbach}
G.~Hadjeres, F.~Pachet, and F.~Nielsen, ``Deep{B}ach: a steerable model for
  bach chorales generation,'' in \emph{International Conference on Machine
  Learning}.\hskip 1em plus 0.5em minus 0.4em\relax PMLR, 2017, pp. 1362--1371.

\bibitem{midinet}
L.~Yang, S.~Chou, and Y.~Yang, ``Midi{N}et: {A} convolutional generative
  adversarial network for symbolic-domain music generation,'' in
  \emph{Proceedings of the 18th International Society for Music Information
  Retrieval Conference, {ISMIR} 2017, Suzhou, China, October 23-27, 2017},
  2017, pp. 324--331.

\bibitem{musicvae}
A.~Roberts, J.~Engel, C.~Raffel, C.~Hawthorne, and D.~Eck, ``A hierarchical
  latent vector model for learning long-term structure in music,'' in
  \emph{International conference on machine learning}.\hskip 1em plus 0.5em
  minus 0.4em\relax PMLR, 2018, pp. 4364--4373.

\bibitem{mt}
C.-Z.~A. Huang, A.~Vaswani, J.~Uszkoreit, I.~Simon, C.~Hawthorne, N.~Shazeer,
  A.~M. Dai, M.~D. Hoffman, M.~Dinculescu, and D.~Eck, ``Music transformer:
  Generating music with long-term structure,'' in \emph{International
  Conference on Learning Representations}, 2018.

\bibitem{popmag}
Y.~Ren, J.~He, X.~Tan, T.~Qin, Z.~Zhao, and T.-Y. Liu, ``Popmag: Pop music
  accompaniment generation,'' in \emph{Proceedings of the 28th ACM
  International Conference on Multimedia}, 2020, pp. 1198--1206.

\bibitem{cwt}
W.-Y. Hsiao, J.-Y. Liu, Y.-C. Yeh, and Y.-H. Yang, ``Compound word transformer:
  Learning to compose full-song music over dynamic directed hypergraphs,'' in
  \emph{Proceedings of the AAAI Conference on Artificial Intelligence}, 2021,
  pp. 178--186.

\bibitem{musebert}
Z.~Wang and G.~Xia, ``Muse{BERT}: Pre-training music representation for music
  understanding and controllable generation,'' in \emph{Proceedings of the 22nd
  International Society for Music Information Retrieval Conference, {ISMIR}
  2021, Online, November 7-12, 2021}, 2021.

\bibitem{popmt}
Y.-S. Huang and Y.-H. Yang, ``Pop music transformer: Beat-based modeling and
  generation of expressive pop piano compositions,'' in \emph{Proceedings of
  the 28th ACM International Conference on Multimedia}, 2020, pp. 1180--1188.

\bibitem{musicbert}
\BIBentryALTinterwordspacing
M.~Zeng, X.~Tan, R.~Wang, Z.~Ju, T.~Qin, and T.~Liu, ``Musicbert: Symbolic
  music understanding with large-scale pre-training,'' in \emph{Findings of the
  Association for Computational Linguistics: {ACL/IJCNLP} 2021, Online Event,
  August 1-6, 2021}, ser. Findings of {ACL}, vol. {ACL/IJCNLP} 2021.\hskip 1em
  plus 0.5em minus 0.4em\relax Association for Computational Linguistics, 2021,
  pp. 791--800. [Online]. Available:
  \url{https://doi.org/10.18653/v1/2021.findings-acl.70}
\BIBentrySTDinterwordspacing

\bibitem{bpe-nlp}
R.~Sennrich, B.~Haddow, and A.~Birch, ``Neural machine translation of rare
  words with subword units,'' in \emph{Proceedings of the 54th Annual Meeting
  of the Association for Computational Linguistics (Volume 1: Long Papers)},
  2016, pp. 1715--1725.

\bibitem{bpe}
P.~Gage, ``A new algorithm for data compression,'' \emph{The C Users Journal},
  vol.~12, no.~2, pp. 23--38, 1994.

\bibitem{arranger}
H.~Dong, C.~Donahue, T.~Berg{-}Kirkpatrick, and J.~J. McAuley, ``Towards
  automatic instrumentation by learning to separate parts in symbolic
  multitrack music,'' in \emph{Proceedings of the 22nd International Society
  for Music Information Retrieval Conference, {ISMIR} 2021, Online, November
  7-12, 2021}, 2021.

\bibitem{musenet}
C.~Payne, ``Muse{N}et,'' \emph{OpenAI Blog}, vol.~3, 2019.

\bibitem{lakhnes}
C.~Donahue, H.~H. Mao, Y.~E. Li, G.~W. Cottrell, and J.~J. McAuley, ``Lakhnes:
  Improving multi-instrumental music generation with cross-domain
  pre-training,'' in \emph{Proceedings of the 20th International Society for
  Music Information Retrieval Conference, {ISMIR} 2019, Delft, The Netherlands,
  November 4-8, 2019}, 2019.

\bibitem{mmm}
J.~Ens and P.~Pasquier, ``Mmm: Exploring conditional multi-track music
  generation with the transformer,'' \emph{arXiv preprint arXiv:2008.06048},
  2020.

\bibitem{musegan}
H.-W. Dong, W.-Y. Hsiao, L.-C. Yang, and Y.-H. Yang, ``Musegan: Multi-track
  sequential generative adversarial networks for symbolic music generation and
  accompaniment,'' in \emph{Thirty-second aaai conference on artificial
  intelligence}, 2018.

\bibitem{pirhdy}
H.~Liang, W.~Lei, P.~Y. Chan, Z.~Yang, M.~Sun, and T.-S. Chua, ``Pirhdy:
  Learning pitch-, rhythm-, and dynamics-aware embeddings for symbolic music,''
  in \emph{Proceedings of the 28th ACM International Conference on Multimedia},
  2020, pp. 574--582.

\bibitem{brown2020language}
T.~B. Brown, B.~Mann, N.~Ryder, M.~Subbiah, J.~Kaplan, P.~Dhariwal,
  A.~Neelakantan, P.~Shyam, G.~Sastry, A.~Askell \emph{et~al.}, ``Language
  models are few-shot learners,'' \emph{arXiv preprint arXiv:2005.14165}, 2020.

\bibitem{vaswani2017attention}
A.~Vaswani, N.~Shazeer, N.~Parmar, J.~Uszkoreit, L.~Jones, A.~N. Gomez,
  {\L}.~Kaiser, and I.~Polosukhin, ``Attention is all you need,'' in
  \emph{Advances in neural information processing systems}, 2017, pp.
  5998--6008.

\bibitem{lineartrans}
A.~Katharopoulos, A.~Vyas, N.~Pappas, and F.~Fleuret, ``Transformers are rnns:
  Fast autoregressive transformers with linear attention,'' in
  \emph{International Conference on Machine Learning}.\hskip 1em plus 0.5em
  minus 0.4em\relax PMLR, 2020, pp. 5156--5165.

\bibitem{adamw}
I.~Loshchilov and F.~Hutter, ``Decoupled weight decay regularization,'' in
  \emph{International Conference on Learning Representations}, 2018.

\bibitem{lakh}
C.~Raffel, ``Learning-based methods for comparing sequences, with applications
  to audio-to-midi alignment and matching,'' Ph.D. dissertation, COLUMBIA
  UNIVERSITY, 2016.

\end{thebibliography}

%
%
%
%
%

\end{document}